%% file: fwave1.tex
\documentstyle[psfig]{europhys}
\input euromacr

\begin{document}
\euro{}{}{}{}
\Date{}
\shorttitle{}
\title{
Possible $f$-wave superconductivity
in Sr$_2$RuO$_4$?}
\author{H. Won\inst{1,2} 
\And K. Maki\inst{2}}
\institute{
\inst{1}Department of Physics, Hallym University,
Chunchon 200-702, South Korea \\
\inst{2}Department of Physics and Astronomy,
 University of Southern California, Los Angeles, CA 90089-0484,
USA}
\rec{}{}
\maketitle
\begin{abstract}
Until recently it has been believed that the
superconductivity in Sr$_2$RuO$_4$
is described by  $p$-wave  pairing.  However, both the
recent specific heat 
and the magnetic penetration depth measurements
on the  purest single
crystals of Sr$_2$RuO$_4$  appear to
 be explained  more consistently in terms of $f$-wave 
superconductivity.
In order to further this hypothesis, we study theoretically 
the thermodynamics and  thermal conductivity of $f$-wave 
superconductors 
in a planar magnetic field. We find the simple expressions
for these quantities when
$H \ll H_{c2} $ and $T \ll T_{c}$, 
which should be readily accessible
experimentally.
\end{abstract}


\section{Introduction}
The recently discovered superconductivity in
Sr$_2$RuO$_4$ has been believed to be 
described in terms of $p$-wave  superconductor with
the full energy gap \cite{maeno,rice}.
For example, the spontaneous spin polarization seen
by muon spin rotation 
experiment \cite{luke} and the flat Knight shift seen by NMR 
\cite{ishida}, 
are consistent with the 
triplet pairing. However, the recent specific heat,
$T_1^{-1}$ in NMR \cite{nishizaki} and the 
superfluid density  of the purest Sr$_2$RuO$_4$ single
crystals with $T_c$ = 1.5 K \cite{bonald} are 
inconsistent with the $p$-wave superconductivity. 
\par 
The $T^2$-dependence of the specific heat, 
the $T^3$-dependence of $T_1^{-1}$,
and the $T$-dependence of
 the superfluid density indicate clearly the presence
of the nodal structure in
the superconducting order parameter.  One possible
interpretation is that the
superconducting order parameter in the $\gamma$-band has the full
gap as assumed earlier, while the ones in the $\alpha$- and
$\beta$-band have the nodal
structure \cite{maki_euro}.
\par
Alternatively, we may consider the possibility
that the superconducting
order parameters in these 3 bands are the same and
described by $f$-wave superconductor 
with the order parameter \cite{yang}
\begin{equation} 
{\bf d}({\bf k}) = \frac{3\sqrt{3}}{2} \Delta 
\hat{d} \hat{k}_3(\hat{k}_1 \pm i \hat{k}_2)^2
\end{equation}
with $\hat{d} \parallel \hat{c}$
which is believed to describe the
superconductivity in UPt$_3$ \cite{heffner}. 
Indeed
the overall temperature dependence of the specific heat
\cite{nishizaki} and 
the superfluid density \cite{bonald}
are described much more consistently by $f$-wave 
superconductor.
We compare
the experimental data of the specific heat \cite{nishizaki}
and the superfluid density \cite{bonald} of
Sr$_2$RuO$_4$ crystals with the theoretical results for 
the weak coupling $p$-wave and $f$-wave
superconductors in Fig.1 and Fig.2, respectively.
Of course, there are still obvious discrepancies
in this identification.
For example, the data for $C_s(T)/\gamma T$ exhibits weakly
convex behavior while the 
theory predicts weakly concave behavior, though the
discrepancy is not so striking.
Also the theory cannot account for the $T^3$-dependence
of the superfluid density
observed in the less pure samples. But this may be due to
the non-locality effect 
suggested by Kosztin and Leggett \cite{kosztin}. 

\begin{figure}
\caption{The specific heat data [5] 
divided by $\gamma T$ where $\gamma$ is Sommerfeld
constant is compared with the theoretical results for 
the isotropic $p$-wave [7]  and
$f$-wave [8] superconductors.
}
\end{figure}

\begin{figure}
\caption{The superfluid density data of single crystal 
Sr$_2$RuO$_4$ [6] is compared with
the theoretical results of $p$-wave [7] 
and $f$-wave [8] superconductors.
}
\end{figure}

\par
In a recent series of papers \cite{maki_physica,maki_m2s}, we have
proposed that the thermal
conductivity tensor in a planar magnetic field  
near $H_{c2}$ 
provide the crucial test of
the symmetry of the underlying superconductivity. We
recall also ingenious thermal conductivity 
experiments \cite{salamon,aubin} have been carried out to elucidate the
nodal structure of $d$-wave
superconductivity in YBCO.
\par
In the following we take the superconducting
order parameter given by 
Eq.(1), and first study the quasi-particle density of
states in the presence of 
both magnetic field and impurities. Starting from the
pioneering work by 
Volovik \cite{volovik},  
we have fully developed technology for this purpose,
at least, for $H \ll H_{c2}$  and 
$T \ll T_c$ \cite{barash,kubert,vekhter,won_cond}.  
The central idea is to introduce the effect of the
magnetic field or the supercurrent in the quasi-particle spectrum through the
Doppler shift\cite{maki_prog}. 
When computing the effect of the magnetic field
to the quasi-particle 
density of states, for example, we take the average of terms
containing the Doppler shift 
over both the quasi-particle momentum and over the
Wigner-Seitz cell containing a single vortex in a real space.
Further impurity scattering is treated in the
unitarity limit as in most
 of unconventional superconductors \cite{yang}. 
\par 
Perhaps the most important result is that the thermal
conductivity tensors exhibit significant
$\tilde{\theta}$-dependence,
where $\tilde{\theta}$ is the angle
between the magnetic field and
the heat current both lying in the $a$-$b$ plane.
Recently the thermal conductivity of
Sr$_2$RuO$_4$ in a planar magnetic field
 has been measured, which does not exhibit
clear $\tilde{\theta}$-dependence 
\cite{tanatar,izawa}.
Of course the $\tilde{\theta}$-dependence
we found for $f$-wave superconductor
is much smaller than the one found
for $p$-wave superconductor \cite{won_prep},
but should be still visible.
Therefore it appears that 
the thermal conductivity data exclude both
$p$- and $f$-wave superconductors from
the candidate for superconductivity in 
Sr$_2$RuO$_4$.
Clearly we have to look for another candidate.
\par
On the other hand, 
the present result should apply to UPt$_3$ in
a low magnetic field with a heat current within 
the $a$-$b$ plane.
We have already studied the thermal conductivity
of $f$-wave superconductor in the vicinity of
$H_{c2}$  \cite{maki_m2s}.
Indeed this calculation reproduces the 
$\tilde{\theta}$-dependence of the thermal conductivity of 
UPt$_3$ observed by Suderow et al \cite{suderow}.

\section{Quasi-particle density of states, the specific
heat, and superfluid density}
In the following we shall use the approach
given in [16]. The residual 
density of states in  $f$-wave  superconductors is given by
\begin{eqnarray}
\frac{N(\omega=0)}{N_0} &=&{\rm Re}\,\,
\langle  \frac{\tilde{\omega}- {\bf v}\cdot {\bf q}}
{\sqrt{(\tilde{\omega} -{\bf v}\cdot {\bf q})^2 - \Delta^2
|f|^2 }}\rangle \big|_{\omega=0}
\nonumber \\
&\simeq&
\frac1{\sqrt{3}}\big\langle
\ln\big(\frac{2}{\sqrt{C_0^2+x^2}}\big) +
x\tan^{-1}(\frac{x}{C_0}) \rangle
\end{eqnarray}
where $f=\frac{3\sqrt{3}}{2}\cos\theta(1-\cos^2\theta)$,
$x=|{\bf v}\cdot {\bf q}|/\Delta$ with $\bf v$ the Fermi velocity, 
${\bf q}$ the superfluid momentum, and 
$C_0 = -i\frac{\tilde{\omega}}{\Delta}\big|_{\omega=0} $
with $\tilde{\omega}$ the renormalized frequency \cite{won_cond}.
In deriving Eq.(2), we
have assumed $C_0,\,\, x \ll 1$.  
Now in  the 
unitarity limit of the impurity scattering
 we obtain \cite{yang}
\begin{equation}
C_0=  \frac{\Gamma}{\Delta}/ 
\big(\frac{N(\omega=0)}{N_0} \big)
\end{equation}
and $\Gamma$ and $\Delta$ are the impurity 
scattering rate and the
superconducting order parameter, respectively.
\par
We can solve Eq.(2) and (3) analytically in the two limiting
cases:
\newline
\par\noindent
a) superclean limit ($C_0 \ll \langle x \rangle \ll 1$, 
i.e. $\displaystyle \frac\Gamma\Delta \ll \frac{H}{H_{c2}} \ll 1$)
\begin{equation}
\frac{N(H)}{N_0} \simeq 
\frac{\pi}{2\sqrt{3}}\langle x\rangle + 
\frac{2}{\pi}
\frac{\Gamma}{\Delta \langle x \rangle} 
\langle \ln(\frac{2}{x}) -1 \rangle
\end{equation}
and
\begin{equation}
C_0 \simeq\frac{2\sqrt{3}}{\pi} 
\frac{\Gamma}{\Delta \langle x \rangle} 
\end{equation}
\par
Finally, following [17] the spatial average
gives
\begin{equation}
\frac{N(H)}{N_0}
\simeq\frac{1}{\sqrt{3}}
\frac{\sqrt{v v^{\prime} e H}} {\Delta}
+ \frac{\Gamma}{\sqrt{v v^{\prime} e H}}
\ln ( \frac{4 \Delta}{\sqrt{v v^{\prime} e H}})
\end{equation}
where $\langle x \rangle \simeq  {\displaystyle
\frac{2}{\pi} \frac{\sqrt{vv^{\prime} eH}}{\Delta}}$
and 
$\displaystyle \langle \ln x \rangle \simeq 
\ln(\frac{\sqrt{vv^{\prime}eH}}{2\Delta}) $
after the  spatial average.
Here $v$ and $v'$ are the Fermi velocity in the $a$-$b$ 
plane and parallel 
to the $c$-axis, respectively.  As shown by Volovik \cite{volovik}
already the density of states
increases like $\sqrt{H}$ for $H\ll H_{c2}$.
\newline
\par\noindent
b) clean limit ($\langle x \rangle \ll C_0 \ll 1$,
i.e.  $\displaystyle \frac{H}{H_{c2}} \ll \frac\Gamma\Delta \ll 1$)
\begin{eqnarray}
\frac{N(H)}{N_0}&\simeq&\frac{N_{\rm imp}}{N_0} \big( 
1+ \frac{\Delta}{2\sqrt{3}\Gamma}
\langle x^2 \rangle  \big)
\nonumber \\
&=&\frac{N_{\rm imp}}{N_0}
(1 + \frac{v v^{\prime} e H}{8\sqrt{3} \Gamma\Delta} 
\ln (\frac{2\Delta}{\sqrt{v v^{\prime} e H} }) )
\end{eqnarray}
and
\begin{equation}
C_0^2\ln (\frac{2}{C_0}) \simeq \sqrt{3}\frac{\Gamma}{\Delta}
-\frac12 \langle x^2 \rangle
\end{equation}
where 
\begin{equation}
\frac{N_{\rm imp}}{N_0}  
\simeq \sqrt{\frac{\Gamma}{2\sqrt{3}\Delta} 
\ln(\frac{4\Delta}{\sqrt{3}\Gamma})}
\end{equation}
Here $\displaystyle \frac{N_{\rm imp}}{N_0}$ 
is the density of states in the $H=0$ case
with  the unitarity  impurity scatterer. 
\par
Also, unlike in $d$-wave superconductors \cite{vekhter,won_cond},
the specific heat
 is independent of the direction of the planar magnetic
field.  Making use of $N/N_ 0$ given in Eqs. (5) and (8) the low temperature
specific heat and the superfluid density are expressed as in
\cite{won_euro} 
\begin{equation}
C_s(T,H) =\frac{2\pi^2}{3} T N(H) 
\end{equation}
and
\begin{equation}
\rho_s(H, T=0) = 1 - N(H)/N_0
\end{equation}

\section{Thermal conductivity tensor}
As already discussed, the angular independence
of the thermal 
conductivity tensor in a planar magnetic field appears to offer
the  test of $f$-wave 
superconductivity. Following the formalism developed by
Ambegaokar-Griffin \cite{griffin},
the thermal conductivity tensor for 
$T \ll \Delta_0$ is given by
\begin{equation}
\kappa_{\parallel}/\kappa_{n}
 =
3 \frac{\Gamma}{\Delta}
\Big\langle
(1-\cos^2\theta) \cos^2\phi
\frac{{\displaystyle\frac12}
\Big(1 + \frac{\displaystyle C_0^2 + x^2 - |f|^2)}
{\displaystyle |(C_0 + ix)^2
+|f|^2|} \Big)}
{{\rm Re}\sqrt{(C_0 + i x)^2 + |f|^2}} \Big \rangle
\end{equation}
and 
\begin{equation}
\kappa_{\perp}/\kappa_n
=
\frac32 \frac{\Gamma}{\Delta}\Big\langle
(1-\cos^2\theta)\sin(2\phi)
\frac{{\displaystyle\frac12}
\Big(1 + \frac{\displaystyle C_0^2 + x^2 - |f|^2)}
{\displaystyle |(C_0 + ix)^2
+|f|^2|}\Big)}
{{\rm Re}\sqrt{(C_0 + i x)^2 + |f|^2)}} \Big \rangle
\end{equation}
and $\kappa_n =\displaystyle
\frac{\pi^2 T n}{6\Gamma m}$, 
the thermal conductivity in the normal state.
\newline
\par\noindent
a) superclean limit($C_0 \ll \langle x \rangle  \ll 1$,i.e.
$\displaystyle \frac{H}{H_{c2}} \ll \frac\Gamma\Delta \ll 1$)
\par
Making use of $C_0$ obtained in Eq.(5) and
integrating over $\cos\theta$ and $\phi$, we obtain
\begin{equation}
\kappa_{\parallel}/\kappa_{n}
\simeq\frac{1}{6} 
\frac{v v^{\prime} eH}{\Delta^2}( 1- \frac13 \cos(2\tilde{\theta}))
\end{equation}
and
\begin{equation}
\kappa_{\perp}/\kappa_n
\simeq
-\frac1{18} \frac{v v^{\prime} eH}{\Delta^2}
\sin(2\tilde{\theta})
\end{equation}
where $\tilde{\theta}$ is the angle between $\bf H$ and $\bf q$
the heat current within the $a$-$b$ plane.
The above expressions may be contrasted with those in
$d$-wave superconductors  which is given  by
\cite{won_cond}
\begin{equation}
\kappa_{\parallel}/\kappa_n \simeq\frac2\pi \frac{v v^{\prime}
eH}{\Delta^2} \big(0.955 + 0.0286\cos(4 \tilde{\theta})\big)^2
\end{equation}
and
\begin{equation}
\kappa_{\perp}/\kappa_n \simeq -\frac2\pi
\frac{v v^{\prime} eH}{\Delta^2} (0.955 + 0.0286\cos(4
\tilde{\theta}))
(0.29\sin(2\tilde{\theta}))
\end{equation}
For $\frac{T}{\Delta} \gg C_0$, $\kappa_{\parallel}$
is given by
\begin{equation}
\kappa_{\parallel}(H=0)=
\frac{3\sqrt{3} \zeta(3) T^2}{2\Delta \sqrt{\sqrt{3}\Gamma
\Delta}} \sqrt{\ln (2 \sqrt{\frac{\Delta}{\sqrt{3}\Gamma}})}
\frac{n}{m}
\end{equation}
and $\zeta(3)=1.202\ldots$.
In this limit the thermal conductivity increase like $T^2$ as in
$d$-wave superconductors \cite{won_cond}.
\newline
\par\noindent
b) clean limit ($\langle x \rangle \ll C_0 \ll 1$,
i.e. $\displaystyle \frac\Gamma\Delta \ll \frac{H}{H_{c2}} \ll 1$)
\begin{eqnarray}
\kappa_{\parallel}/\kappa_0 &\simeq & 1 + \frac1{3} \frac{\langle
(1+\cos(2\phi))x^2 \rangle}{C_0^2} \nonumber \\
&=& 1+ \frac1{12\sqrt{3}}
(1-\frac12\cos(2\tilde{\theta)})
\frac{v v^{\prime} eH}{\Gamma \Delta}
\ln(2\sqrt{\frac{\Delta}{\sqrt{3}\Gamma}})
\ln(\frac{2\Delta}{\sqrt{v v^{\prime} eH}})
\end{eqnarray}
\begin{equation}
\kappa_{\perp}/\kappa_0 \simeq -\frac{1}{24\sqrt{3}}
\sin(2 \tilde{\theta}) 
\frac{v v^{\prime} e H}{\Gamma\Delta}
\ln (2 \sqrt{\frac{\Delta}{\sqrt{3} \Gamma}}) 
\ln (\frac{2\Delta}{\sqrt{v v^{\prime} eH}})
\end{equation}
where $\kappa_0=\displaystyle \frac{\pi^2 T n}
{6\sqrt{3}\Delta m}$ is Lee's universal thermal conductivity \cite{lee}.
Therefore $\kappa_{\perp} \sim -\sin(2\tilde{\theta)}$
appears to be the universal behavior for
$p$-wave, $d$-wave, and
$f$-wave suprconductors.
\par
For ${\bf H} \parallel {\bf c}$, we can derive the corresponding
expressions readily, though
we don't expect any angular dependence.
The quasi-particle density of states is given by
for the superclean limit,
\begin{equation}
N(H)/N_0 \simeq \frac{\pi}{2\sqrt{3}} \frac{v \sqrt{eH}}{\Delta^2}
+ \frac{2}{\pi}
\frac{\Gamma}{v \sqrt{e H}} \ln (\frac{4 \Delta}{v \sqrt{eH}})
\end{equation}
for the superclean limit, and 
\begin{equation}
N(H)/N_0 \simeq \frac{N_{\rm imp}}{N_0} 
( 1 + \frac{v^2 e H}{4 \sqrt{3} \Gamma\Delta}
\ln (\frac{2 \Delta}{v \sqrt{eH}}))
\end{equation}
for the clean limit.
Also the thermal conductivity tensor is given by 
\begin{equation}
\kappa_{\parallel}/\kappa_n \simeq \frac{\pi^2}{24}
\frac{v^2 eH}{\Delta^2}
\end{equation}
for the superclean limit, and
\begin{equation}
\kappa_{\parallel}/\kappa_0
\simeq
1 + \frac{1}{6 \sqrt{3}}
\frac{v^2 eH}{\Gamma\Delta} \ln (2
\sqrt{\frac{\Delta}{\sqrt{3}\Gamma}})
\ln (\frac{2 \Delta}{v\sqrt{eH}}) 
\end{equation}
for the clean limit.
Finally, off-diagonal thermal conductivity has the simple
relation like  
$\kappa_{\perp} = \kappa_{\parallel}
(eB/m)\Gamma_H$ 
where $\Gamma_H$ is the scattering rate related to the Hall
coefficient.

\section{Conclusions}
We have studied theoretically the specific heat
and the thermal 
conductivity tensor in the vortex state of $f$-wave 
superconductivity in a planar magnetic
 field when $H \ll H_{c2}$ and
$T \ll T_{c}$.  The quasi-particle relaxation is assumed to be
due to the impurity scattering in the unitarity limit.
\par
We find for $H\ll H_{c2}$ and $T \ll T_c$ appreciable
$\tilde{\theta}$-dependence for both the diagonal and the 
off-diagonal component of the planar thermal 
conductivity tensor in
a planar magnetic field.
Although the present $\tilde{\theta}$-dependences are much smaller than
those expected for
$p$-wave superconductors,
it is not certain if they are 
consistent with the thermal conductivity data
of Sr$_2$RuO$_4$ crystals.
Therefore, further works on the thermal conductivity tensor in
$f$-wave superconductors and other unconventional
superconductors are highly desirable.
On the other hand, the present result should apply
for UPt$_3$ in phase B in a low magnetic field.

\section{Acknowledgments}
We thank I. Bonalde and Y. Maeno for providing us  the digitized
version of their figure, which we used in construction of Fig.1
and Fig.2.
We are also benefited from discussions with K.
Izawa, T. Ishiguro, Y. Maeno,
Y. Matsuda and M.A. Tanatar on their experimental data
of thermal conductivity in Sr$_2$RuO$_4$.  
HW acknowledges the support from the Korea Research Foundation
under the Professor Dispatching Scheme.
Also HW thanks Dept. of Physics and Astronomy, USC
for their hospitality during her stay.

\begin{minipage}[t]{.40\linewidth}
\vskip 8cm
\centering
\psfig{figure=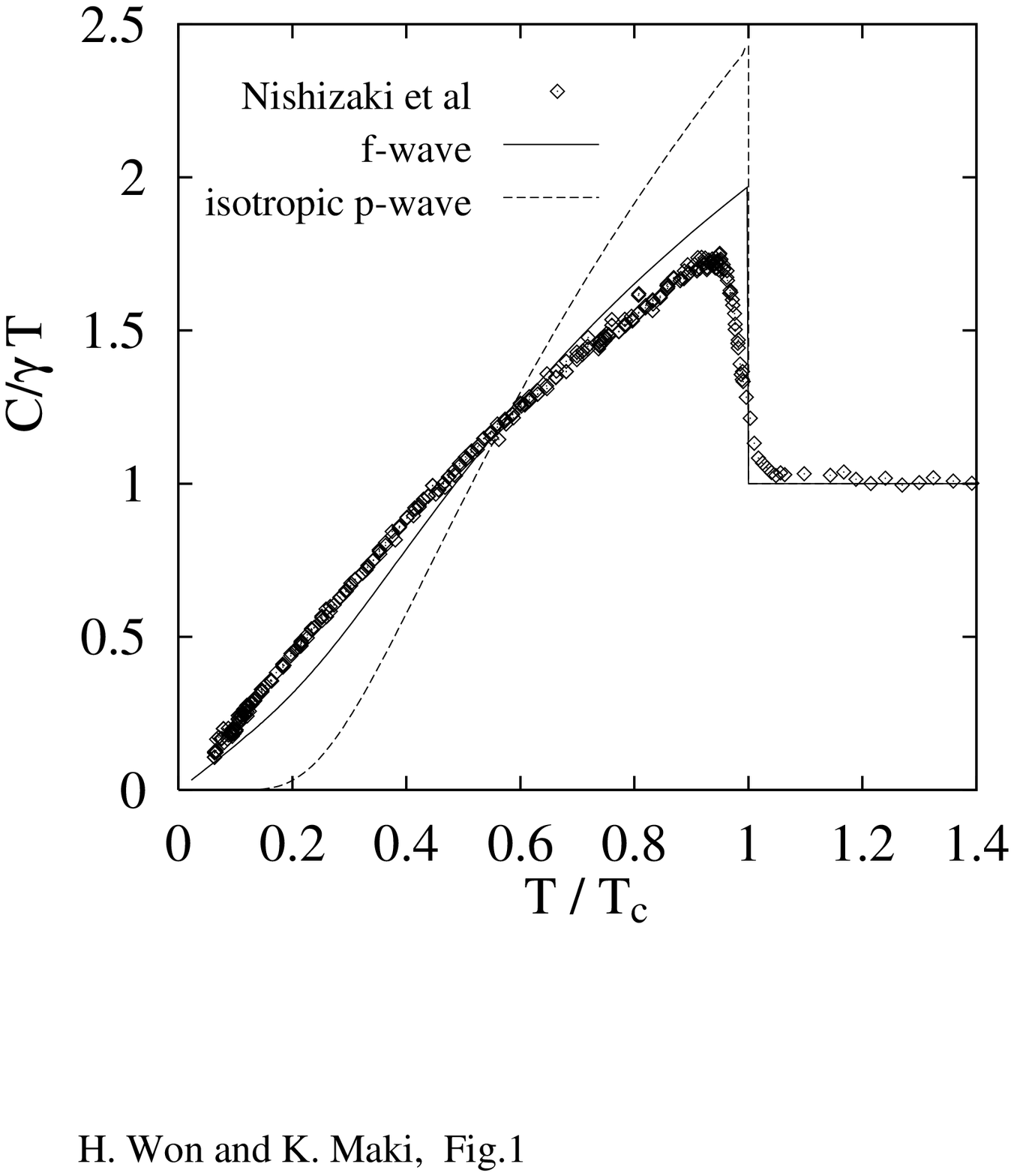}
\end{minipage}

\begin{minipage}[t]{.40\linewidth}
\vskip 8cm
\centering
\psfig{figure=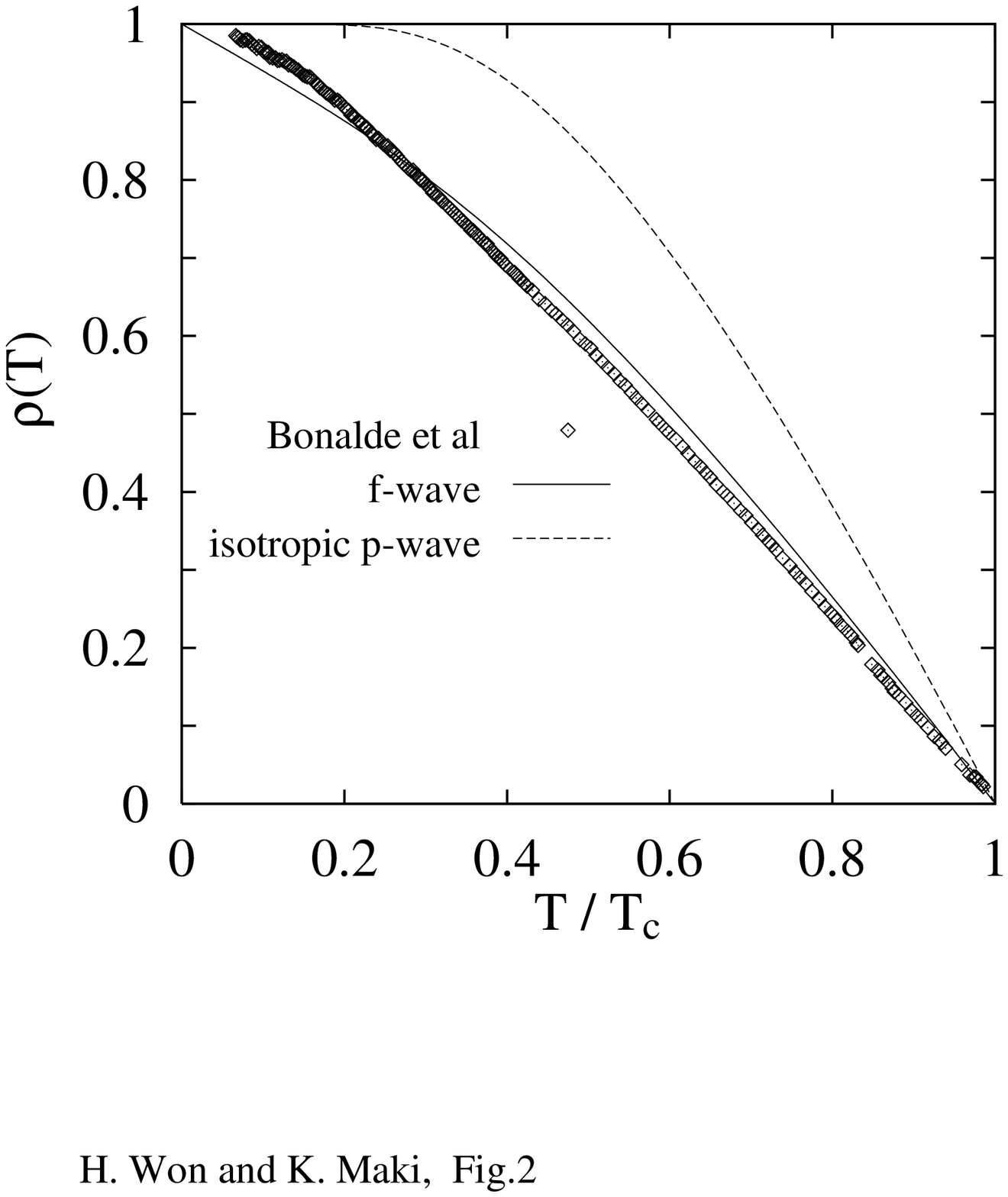}
\end{minipage}

\end{document}

%% file: euromacr.tex
\def\And{{\rm and\ }}

\newif\ifboo \boofalse


%% file: fwave1.bbl
\begin{thebibliography}{99}

\bibitem{maeno}
Maeno Y. et al, Nature {bf 372} (1994) 532; Maeno
Y., Physica C {\bf 282-287} (1997) 206.
\bibitem{rice}
Rice T.M., and Sigrist M., J. Phys. Cond.
Matters {\bf 7} (1995) L643; Sigrist
 M. et al, Physica C {\bf 317-318} (1999) 134.
\bibitem{luke}
Luke G. et al, Nature {\bf 394} (1998) 558.
\bibitem{ishida}
Ishida K. et al, Nature {\bf396} (1998) 658.
\bibitem{nishizaki}
Nishizaki S., Maeno Y. and Mao Z.Q., J. Phys.
Soc. Jpn. {\bf 69}  (2000) 572.
\bibitem{bonald}
Bonalde I., Yanoff B.D., Salamon M.B., Van
Harlingen D.J., Chia E.M.E.,
Mao Z.Q. and Maeno Y., preprint.
\bibitem{maki_euro}
Maki K. and Puchkaryov E., Europhys Lett.
{\bf 50} (2000) 533. 
\bibitem{yang}
Maki K. and Yang G., Fizika {\bf 8} (1999) 345. 
\bibitem{heffner}
Heffner R.H. and Norman M.R., comments on Cond.
Matter Phys. {\bf 17} (1996) 361.
\bibitem{kosztin}
Kosztin I. and Leggett A., Phys.Rev. Lett.
{\bf 79} (1997) 135.
\bibitem{maki_physica}
Maki K., and Won H., Physica C (in press).
\bibitem{maki_m2s}
Maki K., Yang G., and Won H., in Proceedings
of M2S-HTSC VI (Houston, Feb. 2000).
\bibitem{salamon}
Salamon M.B., Yu F., and Kopylor V.N., J.
Superconductivity {\bf  8} (1995) 44
: Yu F., Salamon M.B., Leggett A.J., Lee W.C. and
Ginsberg D.M., Phys. Rev. Lett. {\bf  74} (1995) 5136.
\bibitem{aubin}
Aubin H., Behnia K., Ribault M., Gagnon R. and
Taillefer L., Phys Rev.
Lett.{\bf   78} (1997) 2624.
\bibitem{volovik}
Volovik G.E., JETP Lett. {\bf  58} (1993) 469; Kopnin
N.B. and Volovik G.E. JETP Lett. {\bf  64} (1996) 690.
\bibitem{barash}
Barash Yu S., Svidzinskii A.A., and Mineev
V.P., JETP Lett.{\bf 65} (1997) 638.
\bibitem{kubert}
K\"{u}bert C. and Hirschfeld P.J., Solid State
Commun. {\bf  105} (1998) 459; Phys. Rev. Lett. {\bf  80} (1998) 4963.
\bibitem{vekhter}
Vekhter I., Carbotte J.P., and Nicol E.J., Phys.
Rev. B {\bf  59} (1999) 1417; 
Vekhter I., Hirschfeld P.J., Carbotte J.P., and Nicol
E.J., Phys Rev. B {\bf  59} (1999) R9023.
\bibitem{won_cond}
Won H. and Maki K., cond-mat/0004105
\bibitem{maki_prog}
Maki K., and Tsuento T., Prog. Theor. Phys. {\bf  27}
(1962) 228.
\bibitem{tanatar}
Tanatar M.A. et al, in Proceedings of  M2S-HTSC
IV (Houston, Feb. 2000);
Tanatar M.A. and Maeno Y. (private communication)
\bibitem{izawa}
Izawa K. et al (private communication).
\bibitem{won_prep}
Won H. and Maki K., in preparation.
\bibitem{suderow}
Suderow H., Aubin H., Behnia K., and Huxley A.,
Phys. Lett. A {\bf  234} (1997) 64.
\bibitem{won_euro}
Won H. and Maki K., Europhys. Lett. {\bf  30}, (1995)
427; Phys. Rev. B {\bf  53} (1996) 5927.
\bibitem{griffin}
Griffin A. and Ambegaokar V., Phys. Rev. {\bf  137}
(1964) A1151.
\bibitem{lee}
Lee P.A., Phys. Rev. Lett. {\bf  71} (1993) 1887.
\end{thebibliography}
